\documentclass[paper,11pt]{article}

\makeatletter
\@addtoreset{equation}{section}
\makeatother

\usepackage[toc,title,page]{appendix}
\usepackage{upgreek}

\usepackage{epsfig,amsfonts}
\usepackage{tipa}
\usepackage{mathtools}
\usepackage{graphicx}
\usepackage{caption}
\usepackage{subcaption}
\usepackage{amsmath}
\usepackage{amsthm,amssymb}
\usepackage{dsfont}
\usepackage{mathrsfs}
\usepackage{hhline}
\usepackage{upgreek}
\usepackage[ruled,vlined]{algorithm2e}
\usepackage[linktocpage=true]{hyperref}
\usepackage[dvipsnames]{xcolor}
\usepackage{musicography}
\usepackage[normalem]{ulem}
\usepackage{cite}

\usepackage{tikz}
    \usetikzlibrary{external}             
    \immediate\write18{mkdir -p pgf-img}
\definecolor{purple_nice}{rgb}{0.4,0.2,0.7}
\definecolor{fuel_blue}{RGB}{42,162,185}
\definecolor{YInMn_blue}{RGB}{46, 80, 144}
\definecolor{ultramarine}{RGB}{63, 0, 255}
\definecolor{KLEIN_blue}{rgb}{0, 0.18, 0.65}
\hypersetup{
    colorlinks=true,
    linkcolor=YInMn_blue,
    citecolor=ultramarine,
    filecolor=fuel_blue,
    urlcolor=KLEIN_blue,
}
\def\RBR#1{\left(#1\right)}
\def\SBR#1{\left[#1\right]}
\def\BBR#1{\left\lbrace#1\right\rbrace}
\def\ABR#1{\left\langle#1\right\rangle}
\def\DinD#1{\frac{d}{d#1}}

\def\ve{\varepsilon}
\def\abs#1{\left\vert#1\right\vert}
\def\twomat#1#2#3#4{\left(\begin{array}[c]{cc}
#1 & #2\\
#3 & #4
\end{array}\right)}

\usepackage{enumitem}

\def\be{\begin{equation}}
\def\ee{\end{equation}}
\def\bea{\begin{eqnarray}}
\def\eea{\end{eqnarray}}

\def\XXint#1#2#3{{\setbox0=\hbox{$#1{#2#3}{\int}$}
    \vcenter{\hbox{$#2#3$}}\kern-.5\wd0}}

\def\beq{\begin{equation}}
\def\eeq{\end{equation}}
\def\beqa{\begin{eqnarray}}
\def\eeqa{\end{eqnarray}}

\begin{document}
\begin{titlepage}
\title{{ Topological Gauging and Double Current Deformations }\\}
\author{Sergei Dubovsky$^{1\musFlat}$, Stefano Negro$^{1\musNatural}$ and Massimo Porrati$^{1\musSharp}$\\[0.3cm]}
\date{\footnotesize{$^1$ Center for Cosmology and Particle Physics, New York University,\\ 726 Broadway, New York, NY 10003, U.S.A.\\[0.3cm] $^{\musFlat}$\texttt{\href{mailto:sd103@nyu.edu}{sd103@nyu.edu},}
$^{\musNatural}$\texttt{\href{mailto:sn3101@nyu.edu}{sn3101@nyu.edu},} $^{\musSharp}$\texttt{\href{mailto:mp9@nyu.edu}{mp9@nyu.edu},}\\}}
\maketitle
\begin{abstract}
We study solvable deformations of two-dimensional quantum field theories driven by a bilinear operator constructed from a pair of conserved $U(1)$ currents $J^a$. We propose a quantum formulation of these deformations, based on the gauging of the corresponding symmetries in a path integral. This formalism leads to an exact dressing of the $S$-matrix of the system, similarly as what happens in the case of a $\mathrm{T}\overline{\mathrm{T}}$ deformation. For conformal theories the deformations under study are expected to be exactly marginal. Still, a peculiar situation might arise when the conserved currents $J^a$ are not well-defined local operators in the original theory. A simple example of this kind of system is provided by rotation currents in a theory of multiple free, massless, non-compact bosons. We verify that, somewhat unexpectedly, such a theory is indeed still conformal after deformation and that it coincides with a TsT transformation of the original system. We then extend our formalism to the case in which the conserved currents are non-Abelian and point out its connection with Deformed T-dual Models and homogeneous Yang-Baxter deformations. In this case as well the deformation is based on a gauging of the symmetries involved and it turns out to be non-trivial only if the symmetry group admits a non-trivial central extension. Finally we apply what we learned by relating the $\mathrm{T}\overline{\mathrm{T}}$ deformation to the central extension of the two-dimensional Poincar\'{e} algebra.
\end{abstract}
\end{titlepage}
\newpage
\tableofcontents
\newpage
\section{Introduction}
\label{sec:intro}
A solvable irrelevant deformation of two-dimensional quantum field theories based on the so-called $\mathrm{T}\overline{\mathrm{T}}$ operator \cite{Zamolodchikov:2004ce}, has attracted a lot of interest recently (see \cite{Jiang:2019epa} for a review and for an extended set of references). Initially, this family of theories was constructed at the level of scattering amplitudes \cite{Dubovsky:2013ira}. An independent operator construction, which also leads to exact results for the deformed finite volume spectrum, has been presented in \cite{Smirnov:2016lqw,Cavaglia:2016oda}. An intriguing property of the $\mathrm{T}\overline{\mathrm{T}}$ deformation is that in spite of describing a UV complete quantum theory (at least in the sense that the scattering amplitudes are defined at all energies) the corresponding high energy behavior is not governed by a conformal invariant UV fixed point. This raises the interesting challenge to identify and explore a larger class of quantum theories exhibiting a $\mathrm{T}\overline{\mathrm{T}}$-like UV behavior. Examples of such theories were considered in \cite{Conti:2019dxg, Hernandez-Chifflet:2019sua, Camilo:2021gro, Cordova:2021fnr}, where two-dimensional integrable quantum field theories were deformed by $\mathrm{T}\overline{\mathrm{T}}$-like operators, constructed from higher spin conserved currents. More generally, however, we do not expect the presence of integrability or solvability of the deformation to be necessary conditions for the definition of these kind of theories. A further indication that such a larger family should indeed exist comes from the observation that worldsheet  theories of  large $N$  confining strings are expected to exhibit  a UV behavior similar to $\mathrm{T}\overline{\mathrm{T}}$-deformed theories \cite{Dubovsky:2018dlk}.

A natural language to identify this larger family of  theories is provided by a path integral formulation. In the $\mathrm{T}\overline{\mathrm{T}}$ case, the deformation arises as a consequence of coupling the original quantum field theory to a certain model of topological gravity \cite{Dubovsky:2017cnj,Cardy:2018sdv,Dubovsky:2018bmo}. The net effect of this coupling is to introduce a dynamical system of clocks and rods into the system and implementing these as physical coordinates yields naturally the dressing formula for the $S$-matrix. At the classical level this fact corresponds to the statement that a $\mathrm{T}\overline{\mathrm{T}}$ deformation can be eliminated to an appropriate, field-dependent change of coordinates \cite{Conti:2018tca} coupling the geometry of the system to its energy-momentum.
In the current paper we will further develop this formalism. In fact, our main focus will not be the $\mathrm{T}\overline{\mathrm{T}}$ deformation itself, but rather a generalization, which can be called a \emph{double current deformation}, where the energy-momentum tensor components are replaced by general conserved currents $J^a$. In the case when $J^a$ are conventional spin $1$ currents corresponding to a internal global symmetries, the deformation shares many properties with the $\mathrm{T}\overline{\mathrm{T}}$ one. In particular, the net effect of this double current deformation consists in a dressing of all the operators by the ``Wilson lines'' corresponding to the currents $J^a$. This dressing is the translation to the present case of the dynamical clocks and rods for the $\mathrm{T}\overline{\mathrm{T}}$ case and it is what makes the deformation exactly solvable at the $S$-matrix level. Classically it amounts to a field-dependent redefinition of the field variables themselves. On the other hand, the double current deformation is somewhat less mysterious than the $\mathrm{T}\overline{\mathrm{T}}$ one, as the resulting deformed theory is a conventional quantum field theory whose UV behavior is governed by a conformal fixed point. In \cite{Anous:2019osb,Aguilera-Damia:2019tpe} a gauge theory description has been proposed for deformations involving both $U(1)$ currents and the energy-momentum tensor. Here we follow the same route, considering the case of a deforming operator constructed from two spin $1$ currents. We extend the formalism to the situation when the global symmetries of the undeformed theory are anomalous and derive the deformed S-matrix.

In the special case when the original theory is conformal, a double current deformation is expected to be exactly marginal for spin $1$ currents and the resulting theory should again be conformal. The concrete technical question which is the main focus of the present paper is to study a peculiar version of this situation, when the currents $J^a$ are not well-defined local operators in the original theory.
The simplest example of a situation like this is provided by currents describing the $O(N)$ rotation symmetry of a system of $N$ non-compact massless bosons.  A priori, one might suspect that some subtlety may invalidate the exact marginality of the double current deformation in this case. Indeed, it would be somewhat surprising if there were a prescription to build new conformal theories  starting from an original one, which was not based on a conventional set of conformal data -- {\it i.e.}, the set of conformal primaries and of the corresponding OPE coefficients. 

To study this issue we consider a specific example, which is a theory of four massless scalar fields, where the two currents used to define a deformation correspond to rotations in two orthogonal planes. Somewhat surprisingly, our study does not reveal any subtlety with the double current deformation in this setup. Moreover we show that the family of conformal theories obtained with a double current deformation coincide with an earlier construction using the TsT transformation \cite{Lunin:2005jy} (see also \cite{Apolo:2019zai} for an earlier discussion of the relation between the TsT and $J\bar{J}$). We explain then that this result becomes less surprising if one treats the theory of free, massless, non-compact bosons as the infinite-level limit of a WZW model. Then, before the limit is taken the current involved in the deformation are well-defined spin 1 primaries, making the absence of subtleties in the decompactification limit less surprising.

Finally, we comment on possible non-Abelian generalizations of the double current deformations. As we will see, a necessary condition for the existence of a consistent deformation is the existence of a non-trivial central extension of the symmetry algebra to which the two currents belong. As is well known, internal global symmetries described by conventional semisimple, non-Abelian algebras allow no such central extension. Consequently they do not admit a direct generalization of our definition of double current deformation. However, for those algebras possessing non-trivial central extension, our definition of double current deformation can be applied to yield a class of models that have appeared earlier in the literature under the name of Deformed T-dual Models \cite{Borsato:2016pas,Borsato:2017qsx} and were shown to be equivalent to homogeneous Yang-Baxter deformations \cite{Klimcik:2002zj,Klimcik:2008eq,Delduc:2013qra,Kawaguchi:2014qwa}. A particularly appealing choice of symmetry is the two-dimensional Poincar\'{e} algebra. The corresponding double current deformation is nothing else but the $\mathrm{T}\overline{\mathrm{T}}$ and our perspective put it in direct correspondence with the existence of a central extension of the two-dimensional Poincar\'e algebra.

\section{Gauge representation of double current deformations}
\label{sec:gauge_rep}
\subsection{Initial data}
Let us consider a generic $2$-dimensional Euclidean Quantum Field Theory, associated to some action $\mathcal{A}_0\SBR\Phi$, where $\Phi$ denotes a collection of arbitrary fundamental fields. The local observables of this theory are expectation values of local operators $\mathcal{O}\RBR x \equiv \mathcal{O}\SBR{\Phi\RBR x}$. We can extract information on the theory $\mathcal{A}_0\SBR\Phi$ by looking at the correlators of these local fields
\begin{equation}
	\ABR{\prod_{j=1}^N \mathcal{O}_{j}\RBR{x_j}}_0 = \frac{1}{\mathcal{Z}_0}\intop \SBR{\mathcal{D}\Phi}\, \prod_{j=1}^N \mathcal{O}_{j}\RBR{x_j} e^{-\mathcal{A}_0\SBR\Phi } \;,
\label{eq:undef_correlators}
\end{equation}
where $\mathcal{Z}_0$ is the partition function of the theory. The only specification we will ask to our theory is that it possesses a $U\RBR1\times U\RBR1$ symmetry, that is to say, we require the existence of two independent commuting $U\RBR1$ currents $J^a\RBR x$ with $a=1,2$. Equivalently, we demand that the action $\mathcal{A}_0\SBR\Phi$ can be minimally coupled to a pair of background gauge fields $B^a\RBR x$ in a gauge invariant way\footnote{Here we are supposing, for simplicity, that both symmetries 
do not exhibit external anomalies and so can be gauged. Later we will see that the construction can be extended to the case in which these anomalies are present.}, so that
\begin{equation}
	\mathcal{A}_0\SBR{e^{iq_a\SBR{\Phi}\alpha^a}\Phi\big\vert B-d\alpha} = \mathcal{A}_0\SBR{\Phi\vert B}\;,
\label{eq:gauge_invariance}
\end{equation}
where $q_a\SBR{\Phi}$ are the charges of the fundamental fields $\Phi$ with respect to the $U\RBR1\times U\RBR1$ symmetry with currents $J^a$. From the minimally coupled action $\mathcal{A}_0\SBR{\Phi\vert B}$ the currents can be defined as the response under small variation of the background fields
\begin{equation}
	J_\mu^a\RBR x = -\frac{\delta}{\delta B^\mu_a\RBR x}\mathcal{A}_0\SBR{\Phi\vert B}\Big\vert_{B^a=0}\;,\qquad \partial^\mu J_\mu^a\RBR x\Big\vert_{\mathrm{on-shell}} = 0\;.
\label{eq:current_from_gauge}
\end{equation}
Each of the local operators $\mathcal{O}_j\RBR x$ is charged under the symmetry with, possibly vanishing, charges $q_j^a$ and transforms as usual
\begin{equation}
	\mathcal{O}_j\RBR x \;\longrightarrow\; U_a\mathcal{O}_j\RBR x U_a^{\dagger} = \mathcal{O}_j\RBR xe^{-i q_j^a \alpha_a}\;,\qquad U = e^{\alpha_a\intop dx\,J_0^a}\;.
\end{equation}
The presence of this symmetry in our theory imposes the charge-neutrality condition on all non-vanishing correlators
\begin{equation}
	\ABR{\prod_{j=1}^N \mathcal{O}_j\RBR{x_j}}_0 \neq 0\;\Longleftrightarrow\;\sum_{j=1}^N q_j^a = 0\;,\;a=1,2\;.
\label{eq:charge_neutrality}
\end{equation}
\subsection{Definition of the deformed theory}
\label{subsec:def_def_theo}
Starting from the data presented above, we introduce a one-parameter family of theories by deforming the correlators \eqref{eq:undef_correlators} as follows
\begin{equation}
	\ABR{\prod_{j=1}^N \mathcal{O}_{j}\RBR{x_j}}_{\lambda} = \frac{1}{\mathcal{Z}_{\lambda}}\intop \SBR{\mathcal{D}\Phi \mathcal{D}X \mathcal{D}B}\, \prod_{j=1}^N \tilde{\mathcal{O}}_{j}\RBR{x_j} e^{-\mathcal{A}_0\SBR{\Phi \vert B} - \frac{i}{4\lambda} \ve_{ab} \intop \RBR{dX^a - B^a} \wedge \RBR{dX^b - B^b} } \;.
\label{eq:def_correlators}
\end{equation}
Here  we promote $B^a$'s into dynamical gauge fields and the last term in the exponent can be thought of as a specific mass term for these fields, which is made gauge invariant by introducing the Stueckelberg fields $X^a$'s.
The latter allow us to build gauge invariant operators
\begin{equation}
	\tilde{\mathcal{O}}_j\RBR{x_j} = e^{i q_{j,a} X^a\RBR{x_j}}\mathcal{O}_j\RBR{x_j}\;,
\end{equation}
corresponding to local operators of  the original $\lambda = 0$ theory, dressed by Wilson lines. The partition function is now given by
\begin{equation}
	\mathcal{Z}_{\lambda} = \intop \SBR{\mathcal{D}\Phi \mathcal{D}X \mathcal{D}B}\,  e^{-\mathcal{A}_0\SBR{\Phi \vert B} - \frac{i}{4\lambda} \ve_{ab} \intop \RBR{dX^a - B^a} \wedge \RBR{dX^b - B^b} }\;.
\label{eq:part_func}
\end{equation}
Note that, in the absence of charged operator insertions, the integral over the Stueckelberg fields takes the form of a $\delta$-function imposing the flatness condition of the one-forms $B^a$
\begin{equation}
	dB^a = \ve^{\mu\nu}\partial_{\mu}B_{\nu}^a = 0\;,
\end{equation}
so that the path integral is localized on flat gauge configurations.

Given that the gauge sector $(B,X)$ does not carry any local propagating degrees of freedom, in principle one may integrate it out, which results in a deformed local 
 $\lambda$-dependent action, defined via the path-integral transform
\begin{equation}
	e^{-\mathcal{A}_{\lambda}\SBR\Phi} = \intop \SBR{\mathcal{D}X\mathcal{D}B}\, e^{-\mathcal{A}_0\SBR{\Phi \vert B} - \frac{i}{4\lambda} \ve_{ab} \intop \RBR{dX^a - B^a} \wedge \RBR{dX^b - B^b} }\;.
\label{eq:path_integral_transform}
\end{equation}
Notice that we are taking the deformation parameter to be real $\lambda\in\mathbb{R}$. The imaginary unit in \eqref{eq:def_correlators} guarantees that the deformed action $\mathcal{A}_\lambda$ is real when Wick-rotated to Minkowski space. The expression \eqref{eq:path_integral_transform} is similar to the one considered in \cite{Dubovsky:2018bmo} for the $\mathrm{T}\overline{\mathrm{T}}$ deformation and in \cite{Aguilera-Damia:2019tpe} for joint $\mathrm{J}\overline{\mathrm{T}}$, $\mathrm{T}\overline{\mathrm{J}}$ and $\mathrm{T}\overline{\mathrm{T}}$ deformations. As we are going to see, this is not just a similarity.
 
 In order to gain an understanding of what kind of theory we are dealing with, let us consider the semi-classical limit of  \eqref{eq:path_integral_transform}. 
First, we  can consistently set $X^a$ to zero by choosing the unitary gauge. So our $\lambda$-dependent action is now given by
\begin{equation}
	e^{-\mathcal{A}_{\lambda}\SBR\Phi} = \intop \SBR{\mathcal{D}B}\, e^{-\mathcal{A}_0\SBR{\Phi \vert B} - \frac{i}{4\lambda} \ve_{ab} \intop B^a \wedge B^b }\;.
\label{eq:path_integral_transform_2}
\end{equation}
In the leading semi-classical approximation the integral over $B^a$ is computed simply by fixing the saddle point
\begin{equation}
	B^a = 2i\lambda \ast \tilde{J}^a\RBR{\lambda}\;,\qquad \tilde{J}^a\RBR{\lambda} = \ve^a_{\phantom{a}b} J^b\RBR{\lambda}\;,
\end{equation}
where we introduced the deformed currents, which are implicitly defined by the following equation
\begin{equation}
	J^a_{\mu}\RBR{x\vert\lambda} = - \frac{\delta}{\delta B_a^{\mu}\RBR{x}}\mathcal{A}_0\SBR{\Phi\vert B}\Big\vert_{B^a = 2 i \lambda \ast \tilde{J}^a\RBR{\lambda}}.
\end{equation}
The deformed action takes the following form
\begin{equation}
	\mathcal{A}_{\lambda}\SBR\Phi = \mathcal{A}_0\SBR{\Phi\Big\vert 2i\lambda \ast \tilde{J}\RBR{\lambda}} - i \lambda \ve_{ab} \intop J^a\RBR{\lambda} \wedge J^b\RBR{\lambda}\;,
\end{equation}
from which it is straightforward to derive the flow equation
\begin{equation}
	\frac{d}{d\lambda}\mathcal{A}_{\lambda}\SBR\Phi = i \ve_{ab} \intop J^a\RBR{\lambda} \wedge J^b\RBR{\lambda}\;.
\end{equation}
that first appeared in \cite{Cardy:2019qao} where it was used as a definition of the so called $J^1\wedge J^2$ deformations.
We see that the action defined by the path integral transform \eqref{eq:path_integral_transform} provides a fully quantum definition of these deformations. As we are going to see in \S \ref{sec:nonab_YB_TTbar}, this setting can be extended beyond the $J^1\wedge J^2$ of \cite{Cardy:2019qao} to include deformations generated by non-Abelian currents and the $\mathrm{T}\overline{\mathrm{T}}$ deformation. We are going to call \emph{double-current deformations} the class of theories defined by the path-integral transform \eqref{eq:path_integral_transform}. We will also refer to the whole procedure described in this section as the \emph{topological gauging} of the $U(1)\times U(1)$ symmetry.

Before moving on, let us note a simple fact about the partition function \eqref{eq:part_func} on a plane. In this geometry, the flatness 
condition imposed by the integration over the Stueckelberg fields means that $B^a$ is exact, which further implies that we can set
\begin{equation}
	B^a = 0\;,
\end{equation}
from which it immediately follows that
\begin{equation}
	\mathcal{Z}_\lambda = \mathcal{Z}_0\;.
\end{equation}
\subsection{The $S$-matrix}
\label{subsec:S_mat}
In order to obtain the $S$-matrix for the double-current deformation, we will follow closely the procedure employed in 
\cite{Dubovsky:2018bmo, Dubovsky:2017cnj} for the $\mathrm{T}\bar{\mathrm{T}}$ deformation. Now, let us consider a typical scattering setting in $(1+1)$D Minkowski space-time, in which we have $N$ incoming particles with momenta $p_j^{\mathrm{in}}$ and charges $q_j^{\mathrm{in},a}$, and $M$ outgoing particles with momenta $p_j^{\mathrm{out}}$ and charges $q_j^{\mathrm{out},a}$. From the definition of the theory, we derive the operator equations
\begin{equation}
\label{Heq}
	dB^a = 0\;,\qquad dX^a = 2\lambda \ve^{ab}\ast J_b\;.
\end{equation}
Note that unlike in the derivation of the flow equation for the action described in the previous section, we are not using the saddle point approximation here. Instead, we are working in the operator formalism now, and 
(\ref{Heq}) are exact operator equations in the Heisenberg picture. 
In this case, instead of a static gauge it is convenient to work in a gauge in which $B^a = 0$. Then we see that the net effect of the double current deformation is simply to dress each operator in the theory with a Wilson line operator
\begin{equation}
	\mathcal{O}_j\RBR{x_j}\;\longrightarrow\;\tilde{\mathcal{O}}_j\RBR{x_j} = e^{i q_j^a X_a\RBR{y,x}} \mathcal{O}_j\RBR{x_j}\;,\quad X_a\RBR{y, x} = 2\lambda\ve_{ab}\intop_{\gamma_{y\rightarrow x}} \ast J^b+C_0\;,
\label{eq:X_def}
\end{equation}
where $\gamma_{y\rightarrow x}$ is a curve connecting some fixed base point $y$ to $x$, and $C_0$ is some constant operator. 
As a consequence of current conservation  $\ast J^a$ is a closed $1$-form and the shape of the curve can be deformed at will.
All the creation operators will likewise be dressed by the corresponding Wilson line. This, as we proceed to show, will produce a charge-dependent phase in front of the $S$-matrix. Let us introduce the following notation, that will be useful momentarily
\begin{equation}
	Q_<^a\RBR{x} = \intop_{\RBR{x^0,-\infty}}^{\RBR{x^0,x^1}} \ast J^a\;,\qquad Q_>^a\RBR{x} = \intop_{\RBR{x^0,x^1}}^{\RBR{x^0,\infty}} \ast J^a\;,
\end{equation}
and
\begin{equation}
	Q^a\RBR{x} = Q_<^a\RBR{x} + Q_>^a\RBR{x}\;.
\end{equation}
 To define $X^a$ operators we choose $\gamma_{y\rightarrow x}$ to lie on a constant time slice, $x^0 = y^0 = \mathrm{const}$  and send $y^1\rightarrow -\infty$, while fixing the integration constant symmetrically with respect to $x^1$. This prescription results in the following expression
\begin{equation}
	X^a\RBR{x} =\lambda \ve^a_{\phantom{a}b}\SBR{ Q_<^b\RBR{x} -  Q_>^b\RBR{x}}\;.
\end{equation}
Intuitively, the operator $Q_{<}^a\RBR{x}$ (resp. $Q_{>}^a\RBR{x}$) measures the total $a$-charge lying to the left (resp. right) of the point $x$ along a constant $x^0$-line. 
However, at intermediate times there is no simple expression for how these operators act on a general scattering state. Nevertheless, the intuitive meaning of $Q_\lessgtr$  becomes 
precise in the asymptotic regions $x^0\to\pm\infty$, where all scattering states turn into collections of freely propagating and far separated particles.
 Let us concentrate on the far past $x^0\rightarrow - \infty$. We label the corresponding in-states according to the rapidities of colliding particles, $\beta_j\geq\beta_{j+1}$.
 Particle positions for in-states are also ordered  $x_j \leq x_{j+1}$. According to \eqref{eq:X_def}, the creation operator $A^{\mathrm{in}\dagger}_j\RBR{\beta_j}$ of each incoming particle will be dressed as
\begin{equation}
	A^{\mathrm{in}\dagger}_j\RBR{\beta_j} = \exp\SBR{ i\lambda \ve_{ab} q^{\mathrm{in},a}_j \RBR{Q^b_{<}\RBR{x_j}-Q^b_>\RBR{x_j}}} a^{\mathrm{in}\dagger}_j\RBR{\beta_j}\;,
\end{equation}
where $a_j^{\mathrm{in}\dagger}\RBR{\beta_j}$ are the creation operators for the in particles in the undeformed theory. The successive, ordered action of the operators $A^{\mathrm{in}\dagger}_j$ on the vacuum produces the in-state of the deformed theory. From the above expression we derive the dressing relation\footnote{Na\"{i}vely, a factor $2$ in the exponent might be expected. However one needs to be careful not to double count the charges. In fact, the dressed operators act on the vacuum by creating an incoming particle with rapidity $\beta_j$ and, contextually, measuring -- in the exponent -- the charges of the particles \emph{that have already been created}.}
\begin{equation}
	\left\vert\BBR{\beta_j},\mathrm{in}\right\rangle_{\lambda} = e^{i\lambda \ve_{ab} \sum_{j<k} q_j^{\mathrm{in},a} q_k^{\mathrm{in},b}} \left\vert\BBR{\beta_i},\mathrm{in}\right\rangle_{0}\;.
\end{equation}
The same steps can be carried on \emph{verbatim} to obtain the dressed out state. Taking the overlap of these states, finally, yields the $S$-matrix for the deformed model,
\begin{equation}
\label{Sdef}
    \hat{S}_{\lambda}\RBR{\BBR{\beta_j^{\mathrm{in}}},\BBR{\beta_j^{\mathrm{out}}}} = e^{i\lambda \ve_{ab} \RBR{ \sum_{j<k} q_j^{\mathrm{in},a} q_k^{\mathrm{in},b} + \sum_{j<k} q_j^{\mathrm{out},a} q_k^{\mathrm{out},b}}} \hat{S}_{0}\RBR{\BBR{\beta_j^{\mathrm{in}}},\BBR{\beta_j^{\mathrm{out}}}}\;.
\end{equation}
As we said in the beginning, this derivation is a straightforward extension of the one presented in \cite{Dubovsky:2018bmo, Dubovsky:2017cnj} for the $\mathrm{T}\overline{\mathrm{T}}$ deformed $S$-matrix. A peculiar property of the $\mathrm{T}\overline{\mathrm{T}}$ case is that the corresponding charges generate space-time translations. Then as a result of the dressing procedure Stueckelberg fields acquire the meaning of dynamical (relational) space-time coordinates, which in turn leads to a non-locality of the $\mathrm{T}\overline{\mathrm{T}}$ deformed theories. For conventional internal symmetries the dressing procedure does not affect the locality properties. 

\subsection{A few words on anomalies}
\label{subsec:anomalies}
Strictly speaking, everything we have said up to here holds only if the $U(1)$ symmetries under consideration do not exhibit external anomalies. If this is not the case, however, we can modify the minimally coupled action $\mathcal{A}_0\SBR{\Phi\big \vert B}$ by adding an appropriate quadratic term that cancels the anomaly. Let us be more precise. Suppose that our theory is such that under a gauge transformation $\RBR{\Phi, B^a}\rightarrow \RBR{e^{i q_a\SBR{\Phi}\alpha^a}\Phi, B^a - d\alpha^a}$ the path integral measure transforms anomalously
\begin{equation}
	\SBR{\mathcal{D}\Phi} e^{-\mathcal{A}_0\SBR{\Phi\vert B}} \; \longrightarrow \; \SBR{\mathcal{D}\Phi} e^{-\mathcal{A}_0\SBR{\Phi\vert B} } e^{C_{ab}\intop \alpha^a dB^b}\;.
\label{eq:anom_transf}
\end{equation}
Then we can simply modify the definition \eqref{eq:path_integral_transform} of the action $\mathcal{A}_{\lambda}\SBR{\Phi}$ as follows
\begin{equation}
	e^{-\mathcal{A}_{\lambda}\SBR{\Phi}} = \intop \SBR{\mathcal{D}X \mathcal{D}B} e^{-\mathcal{A}_0\SBR{\Phi\vert B} + C_{ab} \intop X^a dB^b  - \frac{i}{4\lambda} \ve_{ab}\intop \RBR{dX^a - B^a}\wedge \RBR{dX^b - B^b}}\;,
\label{eq:corrected_for_anomaly}
\end{equation}
so that the integrand is properly gauge invariant. We can then proceed with the manipulations we presented above just as in the anomaly-free case. The only difference arises in the the equation for the Stueckelberg fields, which now reads
\begin{equation}
	\SBR{C_{ab} - \frac{i}{2\lambda}\ve_{ab} } dB^b = 0\;.
\label{eq:flatness_w_anomaly}
\end{equation}
The flatness condition $dB^a = 0$ is correctly imposed by this equation, except for the isolated points solving the quadratic equation
\begin{equation}
	4d\lambda^2 + 2it\lambda - 1 = 0\;,\qquad t = \textrm{Tr}\RBR{\ve C}\;,\qquad d = \textrm{Det}\,C\;.
\end{equation}
Similarly, at these points the $B^a$ equations become degenerate and do not allow to solve for the Stueckelberg fields.

So we see that in the case of anomalous symmetries there might exists values of $\lambda$ for which the double-current deformation cannot be defined. These points should have some physical significance. In order to understand this, let us study a simple example: a free compact boson\footnote{We thank Yifan Wang for suggesting us to look at this example.} with action
\begin{equation}
	\mathcal{A}_0\SBR{\varphi} = \intop d^2x\,\frac{\beta^2}{2}\partial_\mu\varphi\partial^\mu\varphi\;,\qquad \varphi \sim \varphi + 2\pi\;.
\end{equation}
This theory possesses two $U(1)$ conserved currents, the \emph{shift} and the \emph{winding} currents
\begin{equation}
	J_{\mathrm{S}} = \beta^2 d\varphi\;,\qquad J_{\mathrm{W}} = \frac{1}{2\pi}\ast d\varphi\;.
\end{equation}
The minimal coupling with the background fields corresponding to these currents has the following form
\begin{equation}
	\mathcal{A}_0\SBR{\varphi\vert B} = \intop d^2x\,\SBR{\frac{\beta^2}{2}\delta^{\mu\nu}\RBR{\partial_\mu\varphi - B_\mu^{\mathrm{S}}}\RBR{\partial_\nu\varphi - B_\nu^{\mathrm{S}}} + i \frac{\varepsilon^{\mu\nu}}{2\pi} \RBR{\partial_{\mu}\varphi - B_\mu^{\mathrm{S}}}B_{\nu}^{\mathrm{W}}}\;.
\label{eq:compact_boson_minimally_coupled}
\end{equation}
This expression can be confirmed by checking that the response under a variation of the background fields correctly yields the currents,
\begin{equation}
	J_{\mathrm{S}} = - \frac{\delta}{\delta B^{\mathrm{S}}}\mathcal{A}_0\SBR{\varphi\vert B}\Big\vert_{B = 0}\;,\qquad J_{\mathrm{W}} = - \frac{\delta}{\delta B^{\mathrm{W}}}\mathcal{A}_0\SBR{\varphi\vert B}\Big\vert_{B = 0}\;.
\end{equation}
However, as it is well known, this system exhibits a mixed anomaly, namely\footnote{This is easily obtained by shifting $B^{\mathrm{W}} \to B^{\mathrm{W}} + d\alpha^{\mathrm{W}}$ in \eqref{eq:compact_boson_minimally_coupled}.}
\begin{equation}
	C_{ab} = 
	\RBR{
		\begin{array}{cc}
			0 & 0 \\
			-\frac{i}{2\pi} & 0
		\end{array}
	}_{ab}\;, \qquad a,b = \mathrm{S},\mathrm{W}\;.
\label{eq:compact_boson_anomaly_matrix}
\end{equation}
Consequently, the deformed model can be defined correctly for all values of $\lambda$ safe for the isolated point $\lambda = \pi$. Importantly, this special value appears as a singularity in the compactification radius of the deformed theory. In fact, if we fix the saddle point in \eqref{eq:corrected_for_anomaly} with \eqref{eq:compact_boson_minimally_coupled} and \eqref{eq:compact_boson_anomaly_matrix}, we find that the double-current deformation amounts to a simple redefinition of the compactification radius $\beta$
\begin{equation}
	\mathcal{A}_\lambda\SBR{\varphi} = \intop d^2x\,\frac{\beta(\lambda)^2}{2}\partial_\mu\varphi\partial^\mu\varphi\;,\qquad \beta\RBR{\lambda} = \frac{\pi}{\vert\lambda - \pi\vert}\beta\;.
\end{equation}
The isolated point $\lambda = \pi$ at which the equations \eqref{eq:flatness_w_anomaly} becomes singular coincide precisely with the singularity of $\beta\RBR{\lambda}$.

Another thing worth noting at this point is that the Stueckelberg field $X$ to be introduced as in \eqref{eq:corrected_for_anomaly} plays the role of the \emph{dual boson} $\tilde{\varphi}$ \cite{TongLN}. From this perspective, the whole deformation procedure described in \S \ref{subsec:def_def_theo} looks very similar to a deformation of the T-duality \cite{Rocek:1991ps}. In fact, it turns out that double-current deformations, in the case of Abelian currents, are nothing else but TsT transformations \cite{Lunin:2005jy}. We will return to this point in \S \ref{subsec:TST}.

\section{Double current deformation of the free complex scalar theory: classical considerations}
\label{sec:classical}

For conventional internal symmetries, physical effects associated with the dressing phase in the deformed S-matrix \eqref{Sdef} are very mild. Indeed, the dressing phase shifts are independent of all kinematical variables, so unlike in the $\mathrm{T}\overline{\mathrm{T}}$ case, there are no scattering time delays  assciated with this dressing. Of course, the double current deformation still changes some of the physical observables, such as the finite volume spectrum. However, given that there is no dimensionful scale associated with these phase shifts, one expects that the deformation should lead to a conformal symmetry if the initial symmetry is itself conformal. Indeed, the operators of the $J\bar{J}$ type are well known to be exactly marginal \cite{Chaudhuri:1988qb} if the corresponding conserved currents are spin one primary operators.

 However, one may worry that something might go wrong when the latter condition is not satisfied for the currents which are used to define the deformation. A simple example of a conserved current which is not a spin one primary (and, in fact, is not a well-defined local CFT operator at all) is provided by the Noether current associated with the phase rotation symmetry of a massless complex scalar boson $\phi$. This current depends on the field $\phi$ itself without any derivatives acting on it.
This leads to the presence of logs in the correlation functions involving this current, and as a result the current does not belong to the set of well-defined local operators in the free boson CFT. 

It will be somewhat surprising if the corresponding double current deformations are still exactly marginal. Indeed, this will mean that it is possible to built new CFTs using ingredients which are not a part of the conventional CFT data. In particular, there are known examples of non-compact $\sigma$-models which are scale invariant, but not conformal invariant \cite{Hull:1985rc}. So one might suspect that the double current deformation will lead to theories like this rather than to proper CFTs in this case.

To test whether these worries are warranted let us study in detail a simple setup of this kind.
Namely, in the next few sections we will focus on a theory of two free, massless complex scalars $\phi^a$, with $a=1,2$, living in $2$-dimensional Euclidean space. Such a theory is described by the action
\begin{equation}
    \mathcal{A}_0\SBR\phi = \intop d^2x\,\SBR{\partial_\mu\phi_a\RBR x}^\dagger\partial^\mu\phi^a\RBR x = \intop \SBR{\ast d\phi_a}^\dagger \wedge \phi^a\;,
\label{eq:free_scalar_action}
\end{equation}
and possesses a $U\RBR1\times U\RBR1$ symmetry
\begin{equation}
	\phi^a\;\longrightarrow\;\phi^a\,e^{-i\alpha_a}\;,
\end{equation}
with associated conserved currents
\begin{equation}
	J_\mu^a\RBR x = i\RBR{\SBR{\phi^a\RBR x}^\dagger \partial_\mu\phi^a\RBR x - \phi^a\RBR x \SBR{\partial_\mu\phi^a\RBR x}^\dagger}\;.
\label{eq:charged_boson_currents}
\end{equation}
We are going to use these two currents to derive the double-current deformed model. To implement the procedure presented in \S\ref{sec:gauge_rep}, we will need the minimally coupled action
\begin{equation}
	\mathcal{A}_0\SBR{\phi\vert A} = \intop d^2x\,\SBR{\RBR{\partial_\mu + iA_\mu^a\RBR x}\phi^a\RBR x}^\dagger \RBR{\partial^\mu + iA_a^\mu\RBR x}\phi_a\RBR x\;,
\end{equation}
from which the currents can be derived by infinitesimally varying the fields $A^a$. The deformed classical action we are looking for can be obtained from
\begin{equation}
	\mathcal{A}_\lambda\SBR{\phi\vert A} = \mathcal{A}_0\SBR{\phi\vert A} + \frac{i}{4\lambda} \ve_{ab} \ve^{\mu\nu} \intop d^2x\,A_\mu^a\RBR x A_\nu^b\RBR x\;,
\label{eq:def_action_gauge}
\end{equation}
after imposing field equations for  $A^a$. As we have proven above, this will be equivalent to the action $\mathcal{A}_\lambda\SBR\phi$ determined by the flow
\begin{equation}
	\DinD\lambda \mathcal{A}_\lambda\SBR\phi = i \ve_{ab}\ve^{\mu\nu} \intop d^2x\,J_\mu^a\RBR x J_\nu^b \RBR x\;.
\label{eq:flow_eq_complex_scalars}
\end{equation}
\subsection{The deformed action}
The form \eqref{eq:def_action_gauge} for the deformed action can be used to easily derive an explicit expression in terms of scalar fields only. All that is needed is the solution to the equation
\begin{equation}
	\frac{i}{2\lambda} \ve_{ab}\ve^{\mu\nu} A_\nu^b = -\frac{\delta}{\delta A_\mu^a} \mathcal{A}_0\SBR{\phi\vert A} = J_a^\mu - 2 A_a^\mu \abs{\phi_a}^2\;,
\end{equation}
where $\abs{\phi_a}^2 = \phi_a^\dagger\phi_a$ and $J_\mu^a$ is as in \eqref{eq:charged_boson_currents}. It is straightforward to find
\begin{equation}
	\begin{split}
		A^1_\mu &= \frac{2 \lambda}{1 + 16 \lambda^2 \abs{\phi^1 \phi^2}^2} \SBR{i \ve_\mu^{\phantom{\mu}\nu} J^2_\nu + 4 \lambda\abs{\phi^2}^2 J_\mu^1}\;,\\
		A^2_\mu &= -\frac{2 \lambda}{1 + 16 \lambda^2 \abs{\phi^1 \phi^2}^2} \SBR{i \ve_\mu^{\phantom{\mu}\nu} J^1_\nu - 4 \lambda\abs{\phi^1}^2 J_\mu^2}\;,
	\end{split}
\end{equation}
from which we derive the action
\begin{equation}
	\mathcal{A}_\lambda\SBR\phi = \intop d^2x\,\SBR{\partial_\mu\phi_a^\dagger \partial^\mu\phi^a + 2\lambda\frac{i \ve^{\mu\nu} J^1_\mu J^2_\mu - 2\lambda\RBR{\abs{\phi^2}^2 J^1_\mu J^\mu_1 + \abs{\phi^1}^2 J^2_\mu J^\mu_2}}{1 + 16\lambda^2 \abs{\phi^1\phi^2}^2}}\;.
\label{eq:def_action_phi}
\end{equation}
By rearranging this expression, we see that this is a non-linear $\sigma$-model with  a non-vanishing $B$-field
\begin{equation}
	\mathcal{A}_\lambda\SBR\phi = \intop d^2x \, \SBR{\delta^{\mu\nu}G_{AB}\RBR\Phi + i \ve^{\mu\nu}B_{AB}\RBR\Phi} \partial_\mu\Phi^A\partial_\nu\Phi^B\;,
\end{equation}
where now $A,B = 1,2,3,4$,
\begin{equation}
	\Phi^{a} = \phi^a\;,\qquad \Phi^{a+2} = \phi^{a\dagger}\;,\qquad a=1,2\;,
\end{equation}
and
\begin{equation}
	G\RBR\Phi = \twomat{g\RBR\Phi}{\gamma\RBR\Phi}{\gamma\RBR\Phi}{g^\dagger\RBR\Phi}\;,\qquad B\RBR\Phi = \twomat{b\RBR\Phi}{\beta\RBR\Phi}{-\beta\RBR\Phi}{b^\dagger\RBR\Phi}\;,
\end{equation}
\begin{equation}
	\begin{split}
		g\RBR\Phi &= \frac{4\lambda^2\phi^{1\dagger} \phi^{2\dagger}}{1+16\lambda^2\abs{\phi^1\phi^2}^2} \twomat{\phi^{1\dagger}\phi^2}{0}{0}{\phi^{2\dagger}\phi^1}\;,\quad
		\gamma\RBR\Phi = \frac{1}{2} \frac{1+8\lambda^2\abs{\phi^1\phi^2}^2}{1+16\lambda^2\abs{\phi^1\phi^2}^2} \mathbb{I}\;,\\
		\beta\RBR\Phi &= \frac{\lambda}{1+16\lambda^2\abs{\phi^1\phi^2}^2} \twomat{0}{\phi^{2\dagger}\phi^1}{-\phi^{1\dagger}\phi^2}{0} \;,\quad
		b\RBR\Phi = - \frac{\lambda\phi^{1\dagger}\phi^{2\dagger}}{1+16\lambda^2\abs{\phi^1\phi^2}^2}\ve\;.
	\end{split}
\end{equation}
We can conveniently use polar coordinates on the target space
\begin{equation}
	\phi^1 = \rho_1 e^{i\theta_1}\;,\qquad \phi^2 = \rho_2 e^{i\theta_2}\;,
\end{equation}
in terms of which the action simplifies greatly
\begin{equation}
	\mathcal{A}_\lambda\SBR{\rho,\theta} = \intop d^2x \, \SBR{\RBR{\partial_\mu \rho_a}^2 + \frac{\RBR{\rho_a\partial_\mu\theta_a}^2}{1 + 16\lambda^2 \rho_1^2 \rho_2^2} - i \frac{8 \lambda \rho_1^2 \rho_2^2}{1 + 16\lambda^2 \rho_1^2 \rho_2^2}\ve^{\mu\nu}\partial_\mu\theta_1\partial_\nu\theta_2}\;.
\label{eq:def_action_polar}
\end{equation}
\subsection{Conserved currents}
Let us return to the expression \eqref{eq:def_action_gauge} of our action, as it turn out to be more convenient. The equations of motion are easily found
\begin{equation}
	\begin{split}
		&\nabla^a_\mu\nabla^{a,\mu}\phi^a = 0\;,\qquad \nabla_\mu^a = \partial_\mu + i A_\mu^a\RBR x\;,\\
		&\frac{1}{2\lambda} \ve^a_{\phantom{a}b} \ve_\mu^{\phantom{\mu}\nu} A_\nu^b = \phi^{a\dagger} \nabla^a_\mu\phi^a - \phi^a \nabla^{a\dagger}_\mu\phi^{a\dagger}\;.
	\end{split}
\end{equation}
As expected from the general arguments, it is not difficult to see that these equations imply flatness of the connections $A^a$
\begin{equation}
	\ve^{\mu\nu}\partial_\mu A^a_\nu = 0\;.
\end{equation}
As usual, it is convenient to adopt complex coordinates
\begin{equation}
	\begin{split}
		z &= x + i y\;,\qquad \partial = \frac{1}{2}\partial_x + \frac{1}{2i} \partial_y \;,\qquad A^a = \frac{1}{2}A^a_x + \frac{1}{2i}A^a_y\;,\\
		\bar{z} &= x - i y\;,\qquad \bar{\partial} = \frac{1}{2}\partial_x - \frac{1}{2i} \partial_y\;,\qquad \bar{A}^a = \frac{1}{2}A^a_x - \frac{1}{2i}A^a_y\;,
	\end{split}
\end{equation}
in terms of which we have
\begin{equation}
	\begin{split}
		\bar{\nabla}^a\nabla^a\phi^a = 0\;,\qquad &\frac{i}{2\lambda} \ve^a_{\phantom{a}b} A^b = \phi^{a\dagger}\nabla^a\phi^a - \phi^a \nabla^{a\dagger}\phi^{a\dagger}\;,\\
		\bar{\partial}A^a = \partial \bar{A}^a\;,\qquad &\frac{i}{2\lambda} \ve^a_{\phantom{a}b} \bar{A}^b = \phi^{a\dagger}\bar{\nabla}^a\phi^a - \phi^a \bar{\nabla}^{a\dagger}\phi^{a\dagger}\;,
	\end{split}
\end{equation}
Note that
\begin{equation}
	\begin{split}
		\nabla^a &= \partial + i A^a\;,\qquad \nabla^{a\dagger} = \partial - i A^a\;,\\
		\bar{\nabla}^a &= \bar\partial + i\bar{A}^a\;,\qquad \bar{\nabla}^{a\dagger} = \bar\partial - i\bar{A}^a\;,
	\end{split}
\end{equation}
and by definition the flatness of $A^a$ implies $\SBR{\nabla^a,\bar{\nabla}^a} = 0$. It is now a matter of elementary algebra to verify that the following quantities
\begin{equation}
	\Psi^a_{(m,n)} = \RBR{\nabla^{a\dagger}}^m\phi^{a\dagger}\RBR{\nabla^a}^n\phi^a\;, \qquad \bar{\Psi}^a_{(m,n)} = \RBR{\bar{\nabla}^{a\dagger}}^m\phi^{a\dagger}\RBR{\bar{\nabla}^a}^n\phi^a\;,
\end{equation}
for $m,n=1,2,\ldots$ are not just conserved, but (anti-)chiral
\begin{equation}
	\bar{\partial}\Psi_{(m,n)}^a = 0\;,\qquad \partial\bar{\Psi}_{(m,n)}^a = 0\;\qquad \forall m,n = 1,2,\ldots\;.
\end{equation}
Limiting to the indices $(m,n) = \RBR{1,1}$, we see that the following two objects
\begin{equation}
	T = \Psi_{(1,1)}^1 + \Psi_{(1,1)}^2\;,\qquad \bar{T} = \bar{\Psi}_{(1,1)}^1 + \bar{\Psi}_{(1,1)}^2\;,
\label{eq:EM_tensor}
\end{equation}
can be identified with the total energy-momentum tensor components of our deformed model. Indeed the same expressions \eqref{eq:EM_tensor} are obtained by Noether procedure from the action \eqref{eq:def_action_gauge}. As each of the summands in \eqref{eq:EM_tensor} is independently conserved and (anti)-chiral, we conclude that our model is classically conformal:
\begin{equation}
	\bar{\partial}T = 0\;,\qquad \partial\bar{T} = 0\;.
\end{equation}

\subsection{Equations of Motion and Deformation map}
The equations of motion can also be extracted from the action \eqref{eq:def_action_gauge}:
\begin{equation}
	\delta^{\mu\nu}\RBR{\partial_\mu + i A_\mu^a}\RBR{\partial_\nu + i A_\nu^a}\phi^a = 0\;.
\end{equation}
It will not serve our purposes to derive explicitly the equations only in terms of the fields $\phi$. What we wish to note is that the above equation can be rewritten as
\begin{equation}
	\partial_\mu\partial^\mu \tilde\phi^a = 0\;,\qquad \tilde\phi^a = e^{i\psi^a}\phi^a\;,\qquad d\psi^a = A^a\;.
\end{equation}
In terms of the fields $\tilde{\phi}$, the stationary point equations for the gauge field take the form
\begin{equation}
	A^a_{\mu} = - 2 i \lambda \ve^{a}_{\phantom{a}b}\ve_{\mu}^{\phantom{\mu}\nu} \tilde{J}^b_{\nu}\;,\qquad \tilde{J}^a_\mu = i\RBR{\tilde{\phi}^{a\dagger}\partial_\mu\tilde\phi^a - \tilde\phi^a\partial_\mu\tilde\phi^{a\dagger}}\;,
\end{equation}
so that
\begin{equation}
	\psi^a\RBR{\gamma_{x_0\rightarrow x}} = -2 i \lambda \ve^a_{\phantom{a}b} \intop_{\gamma_{x_0\rightarrow x}} dy^\mu \ve_\mu^{\phantom{\mu}\nu} \tilde{J}_\nu^b\RBR{y}\;,
\end{equation}
where $x_0$ is an arbitrarily chosen point. This is precisely what we expected from the discussion in \S\ref{sec:gauge_rep}.
\subsection{The double current deformation as a TsT transformation}
\label{subsec:TST}
In \S\ref{subsec:anomalies} we have noticed how the double current deformation procedure described in \S\ref{sec:gauge_rep} resembles a deformation of the T-duality for the case of a compact boson. In fact, it corresponds exactly to the TsT transformation first introduced in \cite{Lunin:2005jy}. The particular case we are studying, the pair of complex scalars, is actually the example presented in \S2 of  \cite{Lunin:2005jy}. Indeed, the action of two free complex scalars \eqref{eq:free_scalar_action}, can be interpreted as a $\sigma$-model on a flat, torsionless $4$-dimensional target space
\begin{equation}
	\mathcal{A}_0 = \intop d^2x\,G_{IJ} \partial_\mu X^I \partial^\mu X^J\;,
\end{equation}
where
\begin{equation}
	G = \left(\begin{array}{c c c c}
		1 & 0 & 0 & 0\\
		0 & \rho_1^2 & 0 & 0\\
		0 & 0 & 1 & 0\\
		0 & 0 & 0 & \rho_2^2
	\end{array}\right)\;, \qquad \left\lbrace\begin{array}{l}
		X^1 = \rho_1 \\
		X^2 = \theta_1 \\
		X^3 = \rho_2 \\
		X^4 = \theta_2
	\end{array}\right.
\end{equation}
and $\phi^1 = \rho_1 e^{i \theta_1}$, $\phi^2 = \rho_2 e^{i \theta_2}$. The target space metric of this model
\begin{equation}
	ds^2 = d\rho_1^2 + d\rho_2^2 + \rho_1^2 d\theta_1^2 + \rho_2^2 d\theta_2^2\;,
\label{eq:flat_target_metric}
\end{equation}
possesses two Abelian isometries $\theta_i\,\rightarrow\,\theta_i + c_i$ with $c_i$ constants. The TsT transformation considered in \cite{Lunin:2005jy} (see also \cite{Hoare:2016wsk}) consists of a $T$-duality transformation $\theta_1\,\rightarrow\,\tilde{\theta}_1$, then a shift $\theta_2\,\rightarrow\,\theta_2 + \gamma \tilde{\theta}_1$ and finally a second $T$-duality $\tilde{\theta}_1\,\rightarrow\,\theta_1$. The resulting background is the following
\begin{equation}
	G = \left(\begin{array}{c c c c}
		1 & 0 & 0 & 0\\
		0 & \frac{\rho_1^2}{1+\gamma^2 \rho_1^2 \rho_2^2} & 0 & 0\\
		0 & 0 & 1 & 0\\
		0 & 0 & 0 & \frac{\rho_2^2}{1+\gamma^2 \rho_1^2 \rho_2^2}
	\end{array}\right)\;,\quad B = \left(\begin{array}{c c c c}
		0 & 0 & 0 & 0\\
		0 & 0 & 0 & \frac{\gamma\rho_1^2\rho_2^2}{1+\gamma^2 \rho_1^2 \rho_2^2}\\
		0 & 0 & 0 & 0\\
		0 & \frac{-\gamma\rho_1^2\rho_2^2}{1+\gamma^2 \rho_1^2 \rho_2^2} & 0 & 0
	\end{array}\right)\;,
\end{equation}
\begin{equation}
	\Phi = -\frac{1}{2} \log\SBR{1+\gamma^2 \rho_1^2 \rho_2^2}\;,
\label{eq:TsT_dilaton}
\end{equation}
on which the $\sigma$-model action
\begin{equation}
	\mathcal{A} = \frac{1}{4\pi\hbar}\intop d^2x\,\sqrt{g}\SBR{\RBR{G_{IJ}g^{\mu\nu} + i B_{IJ}\ve^{\mu\nu}}\partial_{\mu}X^I \partial_{\nu}X^J + \hbar R^{\RBR2}\Phi}\;,
\end{equation}
corresponds exactly to \eqref{eq:def_action_polar} in the flat space-time limit $g\,\rightarrow\,\delta$, at order $\mathcal{O}\RBR{1/\hbar}$ and provided one fixes $\gamma = -4\lambda$.

The fact that the model \eqref{eq:def_action_polar} is equivalent to a free theory \eqref{eq:free_scalar_action} under a TsT transformation, indicates that its invariance under conformal transformations should survive quantization, without suffering from any anomaly. Indeed it was shown in a number of works \cite{Tseytlin:1991wr, Panvel:1992he, Haagensen:1997er, Kaloper:1997ux, Jack:1999av, Parsons:1999ze, Borsato:2020bqo} that appropriate quantum corrections to the $T$-duality exist, up to three loops, that preserve the conformal invariance of the model. In the next section we are going to support this assertion by computing the Weyl anomaly coefficient of the model \eqref{eq:def_action_polar} up to $2$-loops. We will see that these can be made to vanish by an appropriate choice of quantum corrections to the classical metric $G$, flux $B$ and dilaton $\Phi$.
\section{Double current deformation of the free complex scalar theory: the Weyl anomaly}
\label{sec:beta_function}
We have seen that our model is classically conformal. Here we wish to verify that this statement survives at the quantum level. Let us then consider the general bosonic $\sigma$-model action in curved space
\begin{equation}
	\mathcal{A} = \frac{1}{4\pi\hbar}\intop d^2x \SBR{\RBR{\sqrt{g} g^{\mu\nu} G_{IJ} + i \ve^{\mu\nu} B_{IJ}} \partial_\mu X^I \partial_\nu X^J + \hbar \sqrt{g} R^{\RBR2} \Phi}\;,
\label{eq:generic_sigma_model_action}
\end{equation}
where $g_{\mu\nu}$ is the world-sheet metric and $R^{\RBR2}$ the corresponding Ricci scalar curvature. 
For convenience, we introduced here the loop counting parameter $\hbar$.
Flat space conformal invariance of the $\sigma$-model is equivalent to the invariance of (\ref{eq:generic_sigma_model_action}) under Weyl transformations 
\begin{equation}
	g_{\mu\nu}\RBR{x} \; \longmapsto \; e^{2\omega\RBR{x}} g_{\mu\nu}\RBR{x}\;.
\end{equation}
This invariance holds up to a trace anomaly.
Under this Weyl transformation, the various quantities contributing to the action vary as follows
\begin{equation}
	\begin{split}
		\delta_{\omega} G_{IJ} &= -\omega \beta_{IJ}^G + \partial_K G_{IJ} \delta_{\omega} X^K\;,\qquad
		\delta_{\omega} \Phi = \omega (c-\beta^{\Phi}) + \partial_K \Phi \delta_{\omega} X^K\;,\\
		\delta_{\omega} B_{IJ} &= -\omega \beta_{IJ}^B + \partial_K B_{IJ} \delta_{\omega} X^K\;,\qquad
		\delta_{\omega} X^I = \omega \hbar F^I\;.
	\end{split}
\end{equation}
Here $\beta_{IJ}^G$, $\beta_{IJ}^B$ and $\beta^{\Phi}$ are, respectively, the $\beta$ functions for the target space metric, $B$-field and dilaton. 
Note that we explicitly separated the central charge contribution $c$ in the dilaton variation, which describes the trace anomaly.
All variations expand in $\hbar$ as
\begin{equation}
	\beta_{ab}^{G,B}\RBR{G,B} = \sum_{k=1}^{\infty} \hbar^k \beta_{ab}^{\RBR{k}G,B}\RBR{G,B}\;,\qquad \beta^{\Phi}\RBR{G,B,\Phi} = \sum_{k=1}^{\infty} \hbar^k \beta^{\RBR{k}\Phi} \RBR{G,B,\Phi}\;.
\label{eq:beta_expansion}
\end{equation}
After some simple manipulation and integration by parts, the variation of the action can be brought to the following form
\begin{equation}
	\begin{split}
		\delta \mathcal{A} = \frac{1}{4\pi\hbar} \intop d^2x\,&\bigg[ \omega \bigg( \sqrt{g} g^{\mu\nu} \RBR{-\beta_{IJ}^G + 2\hbar \nabla_I F_J} + \\
		&+ \ve^{\mu\nu} \RBR{-\beta_{IJ}^B + \hbar H_{IJ}^{\phantom{IJ}K} F_K }\bigg)\partial_{\mu} X^I \partial_{\nu} X^J +\\
		&+ \omega \frac{\hbar}{2} \sqrt{g} R^{\RBR2} \RBR{c-\beta^{\Phi} + \hbar \partial_K F^K} +\\
		&+2\hbar \bigg(\sqrt{g} g^{\mu\nu} F_K \partial_{\mu} X^K \partial_{\nu} \omega - \sqrt{g} \Delta\omega \Phi \bigg) \bigg]\;,
	\end{split}
\label{eq:sigma_action_variation}
\end{equation}
where $\nabla_I$ denotes the standard covariant derivative on the target space with metric $G_{IJ}$, acting on vector fields as
\begin{equation}
	\nabla_I v^K = \partial_I v^K + \Gamma^K_{\phantom{J}IJ} v^J\;,\qquad \Gamma^K_{\phantom{K}IJ} = \frac{1}{2}G^{KL}\RBR{\partial_I G_{LJ} + \partial_J G_{IL} - \partial_L G_{IJ}}
\end{equation}
and $H_{IJK}$ is the torsion
\begin{equation}
	H_{IJK} = \partial_I B_{JK} + \partial_J B_{KI} + \partial_K B_{IJ}\;.
\end{equation}
For scale transformations $\omega=constant$ so the last two terms in eq.~\eqref{eq:sigma_action_variation} cancel, hence the 
condition for scale invariance generalizes to all loops and nonzero antisymmetric tensor $B_{IJ}$ the results of~\cite{fried1,fried2} 
and~\cite{Hull:1985rc}. 

For general conformal transformations, however, $\omega$ is not a constant. Then, in order for the last two terms in~\eqref{eq:sigma_action_variation} to cancel, one has to choose  the vector $F_I$ controlling the transformation law of the world-sheet scalars to be a gradient~\cite{polch}
\begin{equation}
	F_I = - \nabla_I \Phi\;.
\end{equation}
Finally, requiring that all variations vanish  -- apart from the central charge contribution which reproduces the Weyl anomaly --
 we arrive at the following equations determining the conformal invariance of the action \eqref{eq:generic_sigma_model_action}
\begin{equation}
	\begin{split}
		&\beta_{IJ}^G + 2\hbar \nabla_I \nabla_J \Phi = 0\;,\\
		&\beta_{IJ}^B + \hbar H_{IJ}^{\phantom{IJ}K} \nabla_K \Phi = 0\;, \\
		&\beta^{\Phi} + \hbar \nabla_K \Phi \nabla^K \Phi = 0\;.
	\end{split}
\label{eq:conformal_equations}
\end{equation}

\subsection{1-loop computations}
The $\beta$ functions for a  $\sigma$ model with $B$-field were computed at one loop long ago \cite{Curtright:1984dz,Braaten:1985is,Callan:1985ia} and can be expressed in term of geometrical quantities on the target space. Let us introduce the \emph{generalized connection} $F^K_{\phantom{K}IJ}$ defined as
\begin{equation}
	F^K_{\phantom{K}IJ} = \Gamma^K_{\phantom{K}IJ} - \frac{1}{2} H^K_{\phantom{K}IJ}\;,
\end{equation}
and the associated covariant derivative $\mathscr{D}_I$, acting on vector fields as
\begin{equation}
	\mathscr{D}_I v^K = \partial_I v^K + F^K_{\phantom{K}IJ} v^J = \nabla_I v^K - \frac{1}{2} H^K_{\phantom{K}IJ} v^J\;.
\end{equation}
Then we can define the \emph{generalized Riemann tensor} as the curvature of this connection
\begin{equation}
	\mathscr{D}_I\mathscr{D}_J \omega_K - \mathscr{D}_J\mathscr{D}_I \omega_K + H_{IJ}^{\phantom{IJ}L}\mathscr{D}_L \omega_K = \mathscr{R}_{IJK}^{\phantom{IJK}L}\omega_L\;.
\end{equation}
Some elementary algebra shows that
\begin{equation}
	\mathscr{R}_{IJKL} = R_{IJKL} + \frac{1}{2}\nabla_L H_{IJK} - \frac{1}{2} \nabla_K H_{IJL} + \frac{1}{4} H_{IK}^{\phantom{IK}M} H_{MLJ} - \frac{1}{4} H_{IL}^{\phantom{IL}M} H_{MKJ}\;, 
\end{equation}
with $R_{IJKL}$ being the usual Riemann tensor. Contracting one or two pair of indices we obtain, respectively, the \emph{generalized Ricci tensor} and \emph{generalized Ricci scalar}
\begin{equation}
	\begin{split}
		&\mathscr{R}_{IJ} = \mathscr{R}^K_{\phantom{K}IKJ} = R_{IJ} - \frac{1}{2}\nabla^K H_{KIJ} - \frac{1}{4}H_{IKL}H_J^{\phantom{J}KL}\;, \\
		&\mathscr{R} = \mathscr{R}^I_{\phantom{I}I} = R - \frac{1}{4} H_{IJK}H^{IJK}\;,
	\end{split}
\end{equation}
where $R_{IJ}$ and $R$ are the usual Ricci tensor and scalar. These two quantities satisfy a modified version of the twice contracted Bianchi identity, which will be of use to us momentarily. Let us recall it here
\begin{equation}
	\mathscr{D}^J\mathscr{R}_{IJ} - \frac{1}{2} \mathscr{D}_I \mathscr{R} + H_I^{\phantom{I}JK}\mathscr{R}_{JK} - \frac{1}{12} \nabla_I H^{JKL} H_{JKL} = 0\;.
\label{eq:Bianchi_identity_w_torsion}
\end{equation}

The $1$-loop $\beta$ functions for $G$ and $B$ are the symmetric and anti-symmetric parts\footnote{Our convention for the (anti-)symmetrization of the indices is not normalized, i.e. $A_{\RBR{ab}} = A_{ab} + A_{ba}$.} of the Ricci tensor \cite{Braaten:1985is}
\begin{equation}
	\beta_{IJ}^{\RBR{1}G} = \frac{1}{2} \mathscr{R}_{\RBR{IJ}}\;,\qquad \beta_{IJ}^{\RBR{1}B} = \frac{1}{2} \mathscr{R}_{\SBR{IJ}}\;.
\end{equation}
The $1$-loop $\beta$ function for the dilaton is instead given by the following expression
\begin{equation}
	\beta^{\RBR1 \Phi} = -\frac{1}{2} \nabla^2\Phi - \frac{1}{24} H_{IJK}H^{IJK}\;.
\end{equation}
Thus the equations \eqref{eq:conformal_equations} for $G$ and $B$, taken at $1$ loop, combine into
\begin{equation}
	\mathscr{R}_{IJ} + 2 \mathscr{D}_I \mathscr{D}_J \Phi = 0 \;.
\label{eq:one_loop_combined}
\end{equation}
Applying the identity \eqref{eq:Bianchi_identity_w_torsion} and after some massaging of the expression, we find
\begin{equation}
	\begin{split}
		&2 \mathscr{D}_I \mathscr{D}_J \Phi \mathscr{D}^J \Phi - \frac{1}{2} \mathscr{D}_I \mathscr{D}^2 \Phi - \frac{1}{12} H^{JKL}\nabla_I H_{JKL} = \\
		= &\mathscr{D}_I \RBR{\nabla_J \Phi\nabla^J \Phi - \frac{1}{2}\nabla^2 \Phi - \frac{1}{24} H_{JKL} H^{JKL}} = 0\;.
	\end{split}
\end{equation}
which is consistent  with the third equation in \eqref{eq:conformal_equations} at $1$-loop 
\begin{equation}
	\nabla_I \Phi \nabla^I \Phi - \frac{1}{2} \nabla^2 \Phi - \frac{1}{24} H_{IJK}H^{IJK} = 0\;.
\end{equation}
In summary we see that the request of conformal invariance at $1$-loop can be compactly expressed as
\begin{equation}
	\mathscr{R}_{IJ} + 2 \mathscr{D}_I\mathscr{D}_J \Phi = 0,
\label{eq:1loop_conformal_eq}
\end{equation}
since the equation for $\Phi$ automatically follow from this thanks to the Bianchi identity \eqref{eq:Bianchi_identity_w_torsion}.

Let us now verify that a function $\Phi$ exists for our double current deformed scalar model such that \eqref{eq:1loop_conformal_eq} is satisfied. We will use the polar-coordinates description \eqref{eq:def_action_polar} of the target space, due to its simplicity. The classical metric and $B$-field look as follows
\begin{equation}
	\begin{split}
		G^{\RBR0} &= \left(
		\begin{array}[c]{c c c c}
			1 & 0 & 0 & 0 \\
			0 & \frac{\rho_1^2}{1+16\lambda^2 \rho_1^2 \rho_2^2} & 0 & 0 \\
			0 & 0 & 1 & 0 \\
			0 & 0 & 0 & \frac{\rho_2^2}{1+16\lambda^2 \rho_1^2 \rho_2^2}
		\end{array}
		\right)\;, \\
		B^{\RBR0} &= \left(
		\begin{array}[c]{c c c c}
			0 & 0 & 0 & 0 \\
			0 & 0 & 0 & \frac{4\lambda \rho_1^2 \rho_2^2}{1+16\lambda^2 \rho_1^2 \rho_2^2} \\
			0 & 0 & 0 & 0 \\
			0 & -\frac{4\lambda \rho_1^2 \rho_2^2}{1+16\lambda^2 \rho_1^2 \rho_2^2} & 0 & 0
		\end{array}
		\right)\;.
	\end{split}
\label{eq:classical_data}
\end{equation}
It is then a matter of simple computations -- easily implemented in Mathematica\textsuperscript{\copyright} -- to find the generalized Ricci tensor, whose explicit expression is given in Appendix \ref{sec:beta_func}. More importantly one can verify that the following profile
\begin{equation}
	 \Phi^{\RBR0} = -\frac{1}{2} \log\SBR{1 + 16 \lambda^2 \rho_1^2 \rho_2^2}\;,
\label{eq:classical_f}
\end{equation}
satisfies \eqref{eq:1loop_conformal_eq}. Note that this expression coincide exactly with the one in \eqref{eq:TsT_dilaton}, which was obtained from the TsT transformation of the free theory. We conclude that our deformed model is indeed conformally invariant at 
$1$-loop, thanks to quantum corrections to the transformation law of the world-sheet scalars. In particular we see that
\begin{equation}
	\begin{split}
		\delta_{\omega} \rho_1 &= -16\lambda^2 \omega \rho_1 \frac{\rho_2^2}{1 + 16 \lambda^2 \rho_1^2 \rho_2^2} \hbar\;, \\
		\delta_{\omega} \rho_2 &= -16\lambda^2 \omega \rho_2 \frac{\rho_1^2}{1 + 16 \lambda^2 \rho_1^2 \rho_2^2} \hbar\;, \\
		\delta_{\omega} \theta_a &= 0\;.
	\end{split}
\end{equation}
\subsection{2 loops}
The $\beta$-functions at $2$ loops can be written in terms of the generalized Riemann tensor and the torsion as \cite{Metsaev:1987zx,Braaten:1985is,Metsaev:1987bc,Hull:1987pc,Zanon:1987pp}
\begin{equation}
	\beta^{\RBR2}_{IJ} = \frac{1}{2}\SBR{\mathscr{R}_{IKLM} \mathscr{R}^{KLM}_{\phantom{KLM}J} - \frac{1}{2} \mathscr{R}_{IKLM} \mathscr{R}^{LMK}_{\phantom{KLM}J} + \frac{1}{2} \mathscr{R}_{KIJL} H^K_{\phantom{K}MN} H^{LMN}}\;.
\label{eq:beta_2_loops}
\end{equation}
This expression corresponds to a particular ``minimal subtraction scheme'' \cite{Metsaev:1987zx,Metsaev:1987bc}. Alternative schemes are related to the one above by a redefinition of the the metric and $B$-field at $1$ loop
\begin{equation}
	\RBR{G+B}_{I J} \;\longrightarrow\;\RBR{G+B}_{I J} + \hbar\SBR{\alpha_1 R_{IJ} + \alpha_2 H_{IKL}H_J^{\phantom{J}KL} + \alpha_3 \mathscr{R}_{IJ}}\;.
\end{equation}
To verify the $2$-loop conformal invariance of our model, we are allowed to introduce $1$-loop corrections to the classical quantities, i.e.
\begin{equation}
	G_{IJ} = G_{IJ}^{\RBR0} + \hbar G_{IJ}^{\RBR1}\;,\quad B_{IJ}= B^{\RBR0}_{IJ} + \hbar B_{IJ}^{\RBR1}\;,\quad \Phi =  \Phi^{\RBR0} + \hbar \Phi^{\RBR{1}}\;,
\end{equation}
with $G^{\RBR0}$, $B^{\RBR0}$ and $\Phi^{\RBR0}$ given in \eqref{eq:classical_data} and \eqref{eq:classical_f}.
Using Mathematica\textsuperscript{\copyright} and the expression \eqref{eq:beta_2_loops}, it is then not too difficult to verify that the identity
\begin{equation}
	\beta_{IJ}+ 2 \hbar \mathscr{D}_I\mathscr{D}_J \RBR{\Phi^{\RBR0} + \hbar \Phi^{\RBR1}} = \mathcal{O}\RBR{\hbar^3}\;,
\end{equation}
is satisfied for the following $1$-loop corrections
\begin{equation}
	\begin{split}
		G_{IJ}^{\RBR1} &=  - 32 \lambda^2 \frac{\rho_1 \rho_2}{\RBR{1 + 16 \lambda^2 \rho_1^2 \rho_2^2}^2}\left(
		\begin{array}{c c c c}
			16 \lambda^2 \rho_1 \rho_2^3 & 0 & -1 & 0 \\
			0 & 0 & 0 & 0\\
			-1 & 0 & 16 \lambda^2 \rho_1^3 \rho_2 & 0\\
			0 & 0 & 0 & 0
		\end{array}
		\right)\;, \\
		B_{IJ}^{\RBR1} & = 0\;, \\
		\Phi^{\RBR1} &= - 8 \lambda^2 \frac{\rho_1^2 + \rho_2^2}{1 + 16 \lambda^2 \rho_1^2 \rho_2^2}\;.
	\end{split}
\label{eq:1_loop_corrections}
\end{equation}
The transformation law for the scalars becomes then
\begin{equation}
	\begin{split}
		\delta_{\omega}\rho_1 &= - 16\lambda^2 \omega \rho_1 \SBR{ \frac{ \rho_2^2}{1 + 16\lambda^2 \rho_1^2 \rho_2^2} \hbar +  \frac{1 - 16 \lambda^2 \rho_2^4}{\RBR{1 + 16\lambda^2 \rho_1^2 \rho_2^2}^2}\hbar^2 }\;, \\
		\delta_{\omega}\rho_2 &= - 16 \lambda^2 \omega \rho_2 \SBR{ \frac{ \rho_1^2}{1 + 16\lambda^2 \rho_1^2 \rho_2^2} \hbar + \frac{1 - 16 \lambda^2 \rho_1^4}{\RBR{1 + 16\lambda^2 \rho_1^2 \rho_2^2}^2}\hbar^2 }\;, \\
		\delta_{\omega}\theta_a &= 0\;.
	\end{split}
\end{equation}

We have verified that the model \eqref{eq:def_action_polar} is conformal up to $2$ loops, provided the appropriate quantum corrections are introduced. This, together with the fact that the model is a TsT transform of the $\sigma$-model with target space metric \eqref{eq:flat_target_metric}, is a very strong indication that the model should indeed be fully conformal at the quantum level. A rigorous proof of this statement would require an all-loops argument which, for the moment, eludes us. In the next section we will follow an alternative road: we will consider the two free complex scalar theory \eqref{eq:free_scalar_action} as a decompactification limit of a WZW model with compact target space. In the latter, the double current deformation is an exactly marginal deformation, implying that the deformed model is conformal invariant at all loops.
\section{An explanation from the WZW model}
\label{sec:WZW}

Let us consider the $\mathfrak{g}_k$ WZW model with action
\begin{equation}
	\begin{split}
		\mathcal{A}_k\SBR{g} &= \frac{k}{16\pi}\intop_{\partial \Sigma} d^2x\,\mathrm{Tr}\SBR{\partial^\mu g\partial_\mu g} + \Gamma_\Sigma\SBR{\hat{g}}\;, \\
		\Gamma_\Sigma\SBR{\hat{g}} &=  - i\frac{k}{24\pi} \intop_{\Sigma} d^3x\, \ve_{\alpha\beta\gamma} \mathrm{Tr}\SBR{\hat{g}^{-1}\partial^{\alpha}\hat{g} \hat{g}^{-1}\partial^{\beta}\hat{g} \hat{g}^{-1}\partial^{\gamma}\hat{g}}\;.
	\end{split}
\label{eq:WZW_action}
\end{equation}
Here $\Sigma$ is a $3$-dimensional space, with boundary $\partial \Sigma$. The fundamental field $g\,:\,\partial\sigma \rightarrow \mathfrak{G}$ is an element of the Lie group $\mathfrak{G}$, whose Lie algebra is denoted by $\mathfrak{g}$. $k$ is the level of $\mathfrak{g}$ and $\hat{g}$ denotes the extension of the fundamental field $g$ to the $3$-dimensional space $\Sigma$. As is well known, this theory enjoys a Ka\v{c}-Moody symmetry $\mathfrak{G}\RBR{z} \times \mathfrak{G}\RBR{\bar{z}}$ with local, conserved and chiral currents
\begin{equation}
	\begin{split}
		J\RBR{z} &= k \partial g\RBR{z,\bar{z}} \, g^{-1}\RBR{z,\bar{z}}\;,\qquad \bar{\partial}J\RBR{z} = 0 \;, \\
		\bar{J}\RBR{\bar{z}} &= -k g^{-1}\RBR{z,\bar{z}} \, \bar{\partial} g\RBR{z,\bar{z}}\;,\quad\; \partial\bar{J}\RBR{\bar{z}} = 0\;,
	\end{split}
\end{equation}
where $z=x^1 + i x^0$ and $\bar{z} = x^1 - i x^0$.

In this theory there is no conceptual issue in performing a double current deformation. Indeed, owing to the compactness of the target space, the conserved currents $J(z)$ and $\bar{J}(z)$ are well-defined operators of the theory of dimension $1$. For our purposes, we suppose that $\mathfrak{g}$ contains two commuting generators $g_1$ and $g_2$, with corresponding currents $J_1$, $\bar{J}_1$, $J_2$ and $\bar{J}_2$. Then the deforming operator will be
\begin{equation}
	2\ve^{\mu\nu} J^1_\mu J^2_\nu = i\ve_{ab}J^a\bar{J}^b\;,
\end{equation}
which is a well defined marginal operator, leading to an exactly marginal deformation and, hence, a $1$-parameter family of conformal theories. We argue that the conformality of the deformed theory survives in the decompactification limit. As we proceed to show, in this limit the double-current deformed WZW model coincides with \eqref{eq:def_action_polar} plus a number of spectator free fields.

Let $t_A$ denote a matrix representation of the algebra $\mathfrak{g}$. Then we can parametrize the group element $t$ with target space coordinates $X^A$ as
\begin{equation}
	g\RBR{z,\bar{z}} = e^{\frac{i}{\sqrt{k}}X^A\RBR{z,\bar{z}}t_A}\;.
\end{equation}
The decompactification limit of the WZW model corresponds to the the large level $k\rightarrow\infty$ limit. We see that

\begin{equation*}
	g^{-1}\partial_\mu g \underset{k\rightarrow\infty}{=} \frac{i}{\sqrt{k}}\partial_\mu X^A t_A - \frac{1}{2k} \partial_\mu X^A X^B \SBR{t_A,t_B} - \frac{i}{6k^{3/2}} \partial_\mu X^A X^B X^C \SBR{\SBR{t_A,t_B},t_C} + \mathsf{O}\RBR{k^{-2}}\;.
\end{equation*}
Using this limiting behavior, we can write down the large $k$ expansion of the conserved currents
\begin{equation}
	\begin{split}
		J^A &\underset{k\rightarrow\infty}{=} i\sqrt{k}\partial X^A + \frac{1}{2} f^A_{\phantom{A}BC} \partial X^B X^C + \frac{i}{6 \sqrt{k}} f^A_{\phantom{A}DE}f^E_{\phantom{E}BC} \partial X^B X^C X^D + \mathsf{O}\RBR{k^{-1}}\;, \\
		\bar{J}^A &\underset{k\rightarrow\infty}{=} - i\sqrt{k}\bar{\partial} X^A + \frac{1}{2} f^A_{\phantom{A}BC} \bar\partial X^B X^C - \frac{i}{6 \sqrt{k}} f^A_{\phantom{A}DE}f^E_{\phantom{E}BC} \bar\partial X^B X^C X^D + \mathsf{O}\RBR{k^{-1}}\;,
	\end{split}
\end{equation}
where $J = J^A t_A$ and $\bar{J} = \bar{J}^A t_A$ and $f_{ABC}$ are the structure constants of $\mathfrak g$.
These expressions provide the following perspective on the origin of the ``non-conformal" conserved currents $J_{nc}$ in a free boson CFT. 
To be specific, let us consider the simplest case of the $SU(2)$ group, when $A=1,2,3$ and $f_{ABC}=i\epsilon_{ABC}$.
At  the leading order in the large $k$ limit each of the currents $J^A/\sqrt{k},\bar{J^A}\sqrt{k}$ approaches a well-defined conserved spin 1 primary current. This way one gets (anti)holomorphic shift currents of free bosons. However, the sum $J^A+\bar{J}^A$,  which in components is equal to
\begin{equation}
\label{trcur}
J^A+\bar{J}^A=i\sqrt{k}\epsilon_\mu^{\;\nu}\partial_\nu X^A+\dots\;,
\end{equation}
at the leading order at $k\to\infty$ reduces to a trivially conserved current. For a compact boson this is the winding current discussed in section~\ref{subsec:anomalies}; in a non-compact case there are no states which carry charge under this current. The ``non-conformal" conserved rotation currents arise then as the next-to-leading terms in (\ref{trcur}).

Let us now consider an $SU(2)\times SU(2)$ WZW model and its double current deformation of the form $\lambda(J^3+\bar{J}^3)\wedge(J^{\tilde{3}}+\bar{J}^{\tilde{3}})$ where $J^A$ and $J^{\tilde{A}}$ refer to currents corresponding to two different $SU(2)$ subgroups. This deformation is exactly marginal so the deformed theory is conformal.
Formally, the $J^A$, $J^{\tilde{A}}$ currents diverge at large $k$, so one may expect that it is necessary to take $\lambda\sim 1/k$ in order to obtain a well-defined deformation in the $k\to\infty $ limit. However, in view of (\ref{trcur}), the charges of all states corresponding to $J^A+\bar{J}^A$ stay finite at $k\to \infty$. Hence, one may take the $k\to \infty$ limit of this deformation at constant $\lambda$, which reduces to the model considered in the previous section (with a couple of extra decoupled free scalars).

We believe that this construction makes it less surprising that double current deformation allows to build new CFTs using ingredients which are not a part of the standard conformal data.
At least in the free boson example considered here this deformation can be thought of as a limiting case of a proper CFT construction. It will be interesting to understand whether this viewpoint applies more generally.
\section{Non-Abelian symmetries, central extensions, Yang-Baxter deformations and $\mathrm{T}\overline{\mathrm{T}}$}
\label{sec:nonab_YB_TTbar}
In this work we have considered $2$-dimensional theories deformed by an antisymmetric bilinear operator constructed from a pair of $U(1)$ conserved currents. We have seen that such deformations are obtained by a ``topological gauging'' of the $U(1)\times U(1)$ symmetry. It is natural to ask if it is possible to generalize the procedure to the situation in which we consider a non-Abelian group $G$. Naively adapting the steps as in \S\ref{sec:gauge_rep} leads us to consider actions of the following type
\begin{equation}
	\mathcal{A}_{\lambda}\SBR{\Phi\vert A} = \mathcal{A}_0\SBR{\Phi\vert A} + \frac{i}{4\lambda} \intop \epsilon^{\alpha\beta}\omega(A_\alpha,A_\beta)
\label{eq:non_ab_deformed_action_gauge1}
\end{equation}
where $\omega$ is an antisymmetric 2-form on the Lie algebra $\mathfrak{g}$ of $G$,
\begin{equation}
	\omega(A_\alpha,A_\beta)=-\omega(A_\beta,A_\alpha)\;.
\end{equation}
Not surprisingly, for an arbitrary choice of $\omega$ this deformation does not seem to exhibit any nice properties. In particular, an attempt to integrate out $A$ and to interpret the deformation in terms of a $J\bar{J}$ flow equation in general fails, because the original symmetries are broken at a non-zero value of the deformation parameter.
Note that this type of deformations was first considered in \cite{Borsato:2016pas,Borsato:2017qsx} (for non-linear $\sigma$ models and at the classical level), where they were called \emph{Deformed T-Dual Models} and shown to be equivalent to \emph{homogeneous Yang-Baxter deformations} \cite{Klimcik:2002zj,Klimcik:2008eq,Delduc:2013qra,Kawaguchi:2014qwa}. There it was pointed out that these deformations acquire nice properties, and in particular preserve classical integrability, when the 2-form  
$\omega$ satisfies the cocycle condition,
\[
\omega([A,B],C)+\omega([C,A],B)+\omega([B,A],C)=0.
\]
It was argued there that in this case the gauge invariant form of the action \eqref{eq:non_ab_deformed_action_gauge1}, which is obtained by introducing the Stueckelberg field $X\in \mathfrak{g}$, takes the following form 
\begin{equation}
	\mathcal{A}_{\lambda}\SBR{\Phi\vert A} = \mathcal{A}_0\SBR{\Phi\vert A} + \frac{i}{2\lambda}\intop \omega(X, F)
	+ \frac{i}{4\lambda}\intop\epsilon^{\alpha\beta}\omega(A_\alpha,A_\beta)
	\;,
\label{eq:non_ab_deformed_action_gauge2}
\end{equation}
where $F$ is the field strength corresponding to $A$.
As is well known, non-trivial $2$-cocycles are in $1$-to-$1$ correspondence with the non-trivial central extensions of the algebra $\mathfrak{g}$. In particular there exist none for semi-simple algebras.

With appropriate modifications, the manipulations presented in \S\ref{sec:gauge_rep} can be applied to the action \eqref{eq:non_ab_deformed_action_gauge2}. In particular, we see that the action $\mathcal{A}_{\lambda}\SBR{\Phi}$ obtained by integrating over $X$ and $A$, satisfies the flow equation
\begin{equation}
	\frac{d}{d\lambda} \mathcal{A}_{\lambda}\SBR{\Phi} = i\epsilon^{\alpha\beta} \intop \mathrm{Tr}\SBR{J_\alpha\RBR{\lambda}  R\RBR{J_\beta\RBR{\lambda}}}\;,
\label{eq:non_ab_flow_eq}
\end{equation}
where we assumed that $\mathfrak{g}$ has an invariant scalar product (trace), so that the 2-form $\omega$ can be dualized to a linear operator $\tilde{\omega}:\mathfrak{g}\to\mathfrak{g}$, such that
\[
\omega(A,B)=\frac{1}{2}\mathrm{Tr} \SBR{A\tilde{\omega}(B)-B\tilde{\omega}(A)},
\]
and also assumed that $\tilde{\omega}$ is invertible so that $R =\tilde{ \omega}^{-1}$ obeys the \emph{classical Yang-Baxter equation}
\begin{equation}
	\SBR{R\RBR{A},R\RBR{B}} = R\RBR{\SBR{R\RBR{A},B} + \SBR{A,R\RBR{B}}}\;,
\end{equation}
as a consequence of the cocycle property.
We see then that the class of homogeneous Yang-Baxter deformations is equivalent to a non-Abelian double-current deformation. It is in fact straightforward, using e.g. the expressions in \cite{Delduc:2013fga}, to prove that the action of these models indeed satisfies the flow equation \eqref{eq:non_ab_flow_eq}.

As a note of caution though, let us point out that in the non-Abelian case the action (\ref{eq:non_ab_deformed_action_gauge2}) appears to be only a linearization in $X$ of the full gauge-invariant action. Indeed, in the non-Abelian Stueckelberg procedure one introduces a Stueckelberg field $U$, which takes values in the group $G$, and replaces the gauge field with its gauge transform,
$
A_\alpha \to A^U_\alpha
$. The action (\ref{eq:non_ab_deformed_action_gauge2})  can be obtained then by writing $U=e^X$ and linearizing in $X$.

While postponing a careful analysis of these models to future works, we wish to conclude by spending some words on a particularly interesting case. Consider a generic relativistic invariant theory $\mathcal{A}_0\SBR{\Phi}$ in $(1+1)$D Minkowski space-time. The generators of the Poincar\'{e} algebra $\mathfrak{iso}(1,1)$ are $\BBR{P_1,P_2,K}$, satisfying the usual commutation relations. This algebra admits a non-trivial central extension $\mathfrak{iso}_{\eta}(1,1)$ with non-vanishing commutators
\begin{equation}
	\SBR{P_a,P_b} = \eta \ve_{ab}\;,\qquad \SBR{K,P_a} = \ve_a^{\phantom{a}b}P_b\;.
\label{eq:Poinc_cent_ext}
\end{equation}
According to what we said above, we can deform the action $\mathcal{A}_0\SBR{\Phi}$ using the currents associated to the momentum operators $P_a$. These are nothing else but the components of the energy-momentum tensor, so we expect such a deformation to correspond to the $\mathrm{T}\overline{\mathrm{T}}$ deformation of $\mathcal{A}_0\SBR{\Phi}$. In fact, it is easy to verify that the action  \eqref{eq:non_ab_deformed_action_gauge2} with the choices $A_{\mu} = e_{\mu}^a P_a + k_{\mu} K$ and $\omega(P_a) = \eta \ve_a^{\phantom{a}b}P_b$, $\omega(K) = 0$, is equivalent to the one considered in \cite{Dubovsky:2018bmo,Dubovsky:2017cnj}. Here we see it arising from the ``topological gauging'' of the Poincar\'{e} symmetry. Adapting the discussion in \S\ref{subsec:S_mat}, we conclude that the action $\mathcal{A}_{\lambda}\SBR{\Phi}$, obtained from \eqref{eq:non_ab_deformed_action_gauge2} by integrating $X$ and $A$ satisfies the $\mathrm{T}\overline{\mathrm{T}}$ flow equation
\begin{equation}
	\frac{d}{d\tilde{\lambda}}\mathcal{A}_{\lambda}\SBR{\Phi} = \ve^{ab}\ve_{\mu\nu} \intop d^2x\, T_a^{\mu}\RBR{\lambda}T_b^{\nu}\RBR{\lambda}\;,\quad T_{a}^{\mu} = \frac{1}{\det e}\frac{\delta}{\delta e_{\mu}^a} \mathcal{A}_{\lambda}\SBR{\Phi\vert A}\;,
\end{equation}
where $\tilde{\lambda} = \lambda/\eta$. Additionally, each creation operator is dressed by the phase
\begin{equation}
	a^{\mathrm{in}\dagger}_j\RBR{\beta_j}\; \longrightarrow\;  e^{2i\tilde{\lambda} \ve^{ab} p_a\intop \ast T_b}\, a^{\mathrm{in}\dagger}_j\RBR{\beta_j}\;,
\end{equation}
which leads precisely to the usual CDD factor of $\mathrm{T}\overline{\mathrm{T}}$-deformed theories 
\begin{equation}
	S_{\lambda}\RBR{\BBR{p_i^{\mathrm{in}}},\BBR{p_i^{\mathrm{out}}}} = e^{i\tilde{\lambda} \sum_{j<k}\RBR{p_j^{\mathrm{in}} \wedge p_k^{\mathrm{in}} + p_j^{\mathrm{out}} \wedge p_k^{\mathrm{out}} }} S_{0}\RBR{\BBR{p_i^{\mathrm{in}}},\BBR{p_i^{\mathrm{out}}}}
\label{eq:TTbar_CDD}
\end{equation}
The interesting thing to notice here is the direct relation  between the $\mathrm{T}\overline{\mathrm{T}}$ deformation and the central extension \eqref{eq:Poinc_cent_ext} of the $2$-dimensional Poincar\'{e} algebra. In particular, this perspective somewhat demystifies an intriguing coincidence between the  $\mathrm{T}\overline{\mathrm{T}}$ CDD factor and the additional phase in the multiplication law in the centrally extended Poincar\'e group,
\[
e^{ip^1_\alpha P^\alpha}e^{ip^2_\beta P^\beta}=e^{i(p^1_\alpha +p^2_\alpha)P^\alpha}e^{i\eta p^1\wedge  p^2}\;.
\]

\section{Conclusions}
To summarize, we studied double current deformation of two dimensional quantum field theories using the path integral description based on the topological gauging of symmetries.
In the case of conventional internal symmetries this method allows to calculate the deformed $S$-matrix. 
It is quite robust, and works even in situations where one might have expected that something could go wrong, \emph{e.g.}
in the presence of anomalies and when a conformal field theory is deformed by a ``non-conformal" conserved current. It also allowed us to shed light on the relation between the $\mathrm{T}\overline{\mathrm{T}}$ deformation and the central extension of the two-dimensional Poincar\'e algebra. Hence the topological gauging procedure 
provides a natural starting point for exploring generalizations of solvable double current deformations. In particular, this can be achieved by considering more general actions for the gauge fields and further exploring the non-Abelian case. In particular, it looks 
natural to study a deformation of a conformal field theory based on the topological gauging of the Virasoro symmetry, which is perhaps
 the best known example of a truly non-Abelian algebra with a central extension. The latter example is also somewhat different from 
 the ones considered here, because the CFT vacuum is not invariant under most of the Virasoro generators. We hope to address these
  and related questions in the future.

\section*{Acknowledgements}
We thank Ofer Aharony, Victor Gorbenko, and Yifan Wang for fruitful discussions. This work is supported in part by the NSF grant PHY-2210349, by the BSF grant 2018068 and by the Simons Collaboration on Confinement and QCD Strings. S.N. wishes to thank the Department of Physics of the Universit\`{a} degli Studi di Torino for its kind hospitality
\begin{appendices}
\section{$\beta$ functions}
\label{sec:beta_func}
Here we give the explicit expressions for the $\beta$ functions of the double current deformed scalar model, at $1$ and $2$ loops.
At one loop we have
\begin{equation*}
	\beta^{\RBR1} =  \frac{32 \lambda^2}{\RBR{1+16\lambda^2\rho_1^2\rho_2^2}^2} \left(
	\begin{array}[c]{c c c c}
		b_{11}^{\RBR1}\RBR{\rho_1,\rho_2} & 0 & b_{13}^{\RBR1}\RBR{\rho_1,\rho_2} & 0 \\
		0 & b_{22}^{\RBR1}\RBR{\rho_1,\rho_2} & 0 & b_{24}^{\RBR1}\RBR{\rho_1,\rho_2} \\
		b_{13}^{\RBR1}\RBR{\rho_1,\rho_2} & 0 & b_{11}^{\RBR1}\RBR{\rho_2,\rho_1} & 0 \\
		0 & -b_{24}^{\RBR1}\RBR{\rho_1,\rho_2} & 0 & b_{22}^{\RBR1}\RBR{\rho_2,\rho_1}
	\end{array}
	\right)
\end{equation*}
where
\begin{equation*}
	\begin{split}
		b_{11}^{\RBR1}\RBR{\rho_1,\rho_2} &= \rho_2 ^2 \left(1 - 16 \lambda ^2 \rho_1 ^2 \rho_2 ^2\right)\;, \qquad
		b_{22}^{\RBR1}\RBR{\rho_1,\rho_2} = \rho_1 ^2 \rho_2 ^2\frac{1 - 16 \lambda ^2 \rho_1 ^4}{1 + 16 \lambda ^2 \rho_1 ^2 \rho_2 ^2}\;, \\
		b_{13}^{\RBR1}\RBR{\rho_1,\rho_2} &= 2 \rho_1  \rho_2\;,\qquad\qquad\qquad\quad\;\; b_{24}^{\RBR1}\RBR{\rho_1,\rho_2} = - 4 \lambda \rho_1 ^2 \rho_2 ^2 \frac{ \rho_1 ^2+\rho_2 ^2}{1 + 16 \lambda ^2 \rho_1 ^2 \rho_2 ^2}\;.
	\end{split}
\end{equation*}
The $\beta$ function at $2$-loops, including the contributions coming from the $1$-loop corrections to the metric $G$ and the scalar $f$, has a the same form
\begin{equation*}
	\beta^{\RBR2} =  \frac{64 \lambda^2}{\RBR{1 + 16\lambda^2\rho_1^2\rho_2^2}^4} \left(
	\begin{array}[c]{c c c c}
		b_{11}^{\RBR2}\RBR{\rho_1,\rho_2} & 0 & b_{13}^{\RBR2}\RBR{\rho_1,\rho_2} & 0 \\
		0 & b_{22}^{\RBR2}\RBR{\rho_1,\rho_2} & 0 & b_{24}^{\RBR2}\RBR{\rho_1,\rho_2} \\
		b_{13}^{\RBR2}\RBR{\rho_1,\rho_2} & 0 & b_{11}^{\RBR2}\RBR{\rho_2,\rho_1} & 0 \\
		0 & -b_{24}^{\RBR2}\RBR{\rho_1,\rho_2} & 0 & b_{22}^{\RBR2}\RBR{\rho_2,\rho_1}
	\end{array}
	\right)
\end{equation*}
where the entries are now given by
\begin{equation*}
	\begin{split}
		b_{11}^{\RBR2}\RBR{\rho_1,\rho_2} &= 1 - 8 \lambda ^2 \rho_2 ^2 \left(4 \rho_1 ^2 + \rho_2^2\right) - 256 \lambda ^4 \rho_1 ^2 \rho_2 ^4 \left(\rho_1 ^2-\rho_2 ^2\right) - 2048 \lambda ^6 \rho_1 ^4 \rho_2 ^8\;, \\
		b_{22}^{\RBR1} \RBR{\rho_1,\rho_2} &= \rho_1 ^2 \frac{1 - 8 \lambda ^2\rho_2 ^2 \left(4 \rho_1 ^2 + \rho_2 ^2\right) + 128 \lambda ^4 \rho_1 ^2 \rho_2 ^2 \left(\rho_1 ^4+2 \rho_1 ^2 \rho_2 ^2+3 \rho_2 ^4\right) - 6144 \lambda ^6 \rho_1 ^8 \rho_2 ^4}{1 + 16 \lambda ^2 \rho_1 ^2 \rho_2 ^2}\;,  \\
		b_{13}^{\RBR1} \RBR{\rho_1,\rho_2} &= - 16 \lambda ^2 \rho_1  \rho_2  \left(\rho_1 ^2+\rho_2 ^2\right) \left(1 - 16 \lambda ^2 \rho_1 ^2 \rho_2 ^2\right)\;, \\
		b_{24}^{\RBR1} \RBR{\rho_1,\rho_2} &= - 8 \lambda  \rho_1 ^2 \rho_2 ^2 \frac{1 - 4 \lambda ^2 \left(\rho_1 ^4+4 \rho_1^2 \rho_2 ^2+\rho_2 ^4\right) + 192 \lambda ^4 \rho_1 ^2 \rho_2 ^2 \left(\rho_1 ^4+\rho_2 ^4\right)}{1 + 16 \lambda ^2 \rho_1 ^2 \rho_2 ^2}
	\end{split}
\end{equation*}
\end{appendices}
\bibliography{Biblio3}
\end{document}